# Coulomb screening and scattering in atomically thin transistors across dimensional crossover

Shihao Ju, Binxi Liang, Jian Zhou, Danfeng Pan, Yi Shi, Songlin Li*

*National Laboratory of Solid-State Microstructures, Collaborative Innovation Center of Advanced Microstructures, and School of Electronic Science and Engineering, Nanjing University, Nanjing, Jiangsu 210023, China*

*Author to whom correspondence should be addressed: sli@nju.edu.cn

**Abstract:** Layered two-dimensional dichalcogenides are potential candidates for post-silicon electronics. Here, we report insightfully experimental and theoretical studies on the fundamental Coulomb screening and scattering effects in these correlated systems, in response to the changes of three crucial Coulomb factors, including electric permittivity, interaction length, and density of Coulomb impurities. We systematically collect and analyze the trends of electron mobility with respect to the above factors, realized by synergic modulations on channel thicknesses and gating modes in dual-gated $MoS_2$ transistors with asymmetric dielectric cleanliness. Strict configurative form factors are developed to capture the subtle parametric changes across dimensional crossover. A full diagram of the carrier scattering mechanisms, in particular on the pronounced Coulomb scattering, is unfolded. Moreover, we clarify the presence of up to 40% discrepancy in mobility by considering the permittivity modification across dimensional crossover. The understanding is useful for exploiting atomically thin body transistors for advanced electronics.

**Keywords:** Two-dimensional materials, field-effect transistors, transition metal dichalcogenides, charged impurities, Coulomb screening, scattering mechanisms

With the silicon microelectronics approaching its physical limit, layered two-dimensional (2D) transition metal dichalcogenides (TMDs) that feature well-defined surface flatness and atomic-scale thickness, such as molybdenum disulfide ($MoS_2$), are considered as promising candidates for constructing atomically thin body field-effect transistors (FETs), to minimize the notorious short channel effect in advanced technological nodes beyond silicon.[1–3] FETs consisting of the 2D TMD channels are reported to exhibit merits including high intrinsic charge mobility ($\mu$)[4–7] and current density,[8] and superior immunity to short channel effect till gate length below one nanometer.[9,10] In practical devices, however, $\mu$ of these extremely thinned channels is highly sensitive to the interfacial charged impurities (CIs) randomly distributed on the surfaces of deposited dielectrics,[11–14] arising mainly from the unsaturated chemical bonds. The interfacial CIs represent a type of long-range scattering mechanism, as they perturb the periodic lattice potentials and pose scattering to channel charges through the long-range Coulomb interaction. When charge carriers are all confined within the atomic scales, their interaction distance to CIs becomes highly reduced and, thus, the scattering effect is greatly enhanced.[13] Hence, it is of both fundamental and technological importance to fully understand the scattering physics of interfacial CIs and the related Coulomb screening response within channels.

In general, the CI scattering rate ($\tau_{CI}^{-1}$) is associated with the screened Coulomb potential within the FET channels of different configurative structures, following the general form of $\tau_{CI}^{-1} \propto n_{CI} V(d)^2 F(\varepsilon_s, d) \varepsilon_{2D}^{-1}(\varepsilon_s, d)$,[11] where $n_{CI}$, $\varepsilon_s$, $d$, $V(d)$, $F(\varepsilon_s, d)$, $\varepsilon_{2D}(\varepsilon_s, d)$ denote the density of CIs, the electric permittivity of semiconductor, the scattering distance, the Coulomb potential, the configurative form factor of device, and the 2D dielectric function, respectively. The last two terms are essential parameters in analyzing the Coulomb scattering events in FETs and depend highly on device structures. In theoretical calculation, they are normally treated in momentum space, with random phase approximation to account for electron correlation.[13–15] Basically, there is no analytical description for them. The situation becomes more complicated when dimensional crossover of channels occurs, which makes $\varepsilon_s$ and self-screening dielectric response vary with channel thickness.[15] As channels evolve into the atomic scales, the lattice atoms undergo subtle relaxations that result in remarkable changes in polarization capability of materials under electromagnetic fields. For instance, even for the static $\varepsilon_s$ of few-layer TMDs, the values remain controversial.[3,16–19] To date, most theoretical works employ simplified treatments, using delta and trigonometric approximation for carrier distribution, for handling $F(\varepsilon_s, d)$ and $\varepsilon_{2D}(\varepsilon_s, d)$, which work well for the monolayer limit but become invalid for thick channels. Moreover, no works have analyzed the impact of $\varepsilon_s$ variation on the actual dielectric screening in TMDs by including the dimensional crossover from the 2D to 3D regime. As such, the previous calculations may cause unacceptable deviation to reflect their actual effects on electronic characteristics.

In this study, we provide an insightfully experimental and theoretical understanding on the realistic effects of Coulomb screening and scattering under the dimensional crossover regimes. Dual-gated FET structures, with lopsided cleanliness (~10 fold) between the top and bottom dielectrics and with varied channel thickness ($t_s$) from monolayer (1L) to pentalayer (5L), operated under different gating modes, including individual bottom or top, and dual gating modes (denoted by BGM, TGM, and DGM, respectively), were employed to broadly modulate the distribution of charge carriers within channels and the effective $d$ between interfacial CIs and carriers at every adopted $t_s$. Such a strategy allows for multiparametric controls on the three crucial Coulomb factors, i.e., $n_{CI}$, $d$, and $\varepsilon_s$, facilitating a direct and thorough evaluation of their impacts on charge screening and carrier scattering. Furthermore, strict $F(\varepsilon_s, d)$ and $\varepsilon_{2D}(\varepsilon_s, d)$ are developed to capture the subtle parametric changes across dimensional crossover. The trends of intrinsic low-temperature $\mu$ versus temperature ($T$) and $t_s$ were systematically collected and analyzed, together with theoretical calculations, to elucidate the consequence on electronic performance of 2D channels under practical conditions.

Figure 1a–d depicts the fabrication process for the cleanliness-asymmetric dual-gated FETs. The stacks of the bottom gate (Al) and dielectric (10-nm Al$_2$O$_3$) are first defined on SiO$_2$/Si substrates (Figure 1a), followed by the transfer of high-quality exfoliated TMD channels (Figure 1b). Afterward, contact electrodes (Figure 1c), top dielectrics (10-nm Al$_2$O$_3$), and top gates (Ni/Au) (Figure 1d) are prepared in sequence to accomplish the entire FET structure. The schematic three-dimensional diagram is shown in Figure 1g. Figure 1e,f shows the optical images for a typical device before and after depositing the TG stacks,

respectively, where the top and bottom gates are separated only by a thin 10 nm $Al_2O_3$ layer. Hence, its quality represents a critical factor in device fabrication and deserves a check. Thus, an atomic force microscope (AFM) is utilized to evaluate the surface quality of the dielectric layers and transferred channels. As shown in Figure 1h–j, their surface roughness values are all within a low range of 0.3–0.5 nm, ensuring excellent electrical insulation. Here both the bottom and top dielectric layers are prepared by atomic layer deposition (ALD) with the same nominal thickness of 10 nm to create a physically symmetric geometry for the encapsulated TMD channels, but ~10-fold asymmetry in surface cleanliness is found, owing to the presence of chemical residues and partial oxidation of seeding layers on the top dielectrics. Nevertheless, the asymmetric cleanliness in channel surfaces provides a unique opportunity for probing the Coulomb screening effect in TMD FETs.

Figure 1k shows the typical transfer curves, i.e. drain-source current ($I_{ds}$) versus gate voltage ($V_g$), for a dual-gated $MoS_2$ FET operated under the three distinctive gating modes, where n-type conduction behavior is consistently observed. As shown in the wiring geometry (Figure 2a,b), during single gating modes (i.e., BGM and TGM) only the working gate is biased at sweeping voltages with keeping the other idle one grounded, to prevent additional capacitive coupling.[20] Hence, the potential of the channel surface near the idle gate is close to the ground, which renders the channel undepleted, leading to an incomplete off state (current $10^{-7}$~$10^{-9}$ A) even under a relatively high $V_g$ of -6 V. By contrast, the channel becomes fully depleted under DGM (off-state current near $10^{-14}$ A), or when the idle gate is floated (Figure S1a,b). The discrepancy in off-state current between the single or dual gating modes is a natural consequence of the distinct electrostatic gating effects.

The reason why the screening effect can be probed via asymmetric gating or through varying $t_s$ lies in the fact that both actions can change at least one of the three critical scattering parameters $n_{CI}$, $\varepsilon_s$, or $d$, which are responsible for the effective Coulomb potential from the interfacial CIs. Figure 2a–f depicts the schematic diagrams for the wiring geometry of the three alternative gating modes (i.e., BGM, TGM, and DGM) and corresponding charge distribution in the channels with two $t_s$s (i.e., 1L and 5L). As gating changes among BGM, TGM, and DGM, the carrier distribution and, thus, the most relevant scattering interface (associated with the parameter $n_{CI}$) and/or distance (i.e., the parameter $d$) would change accordingly. Likewise, when varying $t_s$ with gating mode fixed, the parameter $d$ would also be finely modulated. This concept can be easily understood when comparing the various diagrams of charge distribution shown in Figure 2.

To rule out the effect of contact resistance and attain the intrinsic electronic behavior, we employed the four-terminal measurements for all devices. Figure 3a–c illustrates the intrinsic channel conductivity ($\sigma$) for a typical 5L MoS$_2$ channel from 8 to 300 K under different gating modes. The channel operates properly under all gating modes because the $T$-varying transfer curves show the expected intersectional points near 40 μS, which are the sign of transition from semiconducting to metallic conduction due to field gating. Under different gating modes, $\sigma$ shows remarkable discrepancies in both magnitude and $T$ dependence. Generally, it follows the trend $\sigma_{DGM} > \sigma_{BGM} > \sigma_{TGM}$ at the same biasing and thermal conditions. At 300 K, the values are 240, 173, and 136 μS under DGM, BGM, and TGM, respectively. Accordingly, at 8 K they rise to 1847, 1070, and 577 μS, respectively. The ratios of $\sigma_{8K}/\sigma_{300K}$ amount to 7.7, 6.2, and 4.2 for each gating mode. These discrepancies become more evident when plotting the 3D contour of channel resistance ($R_s$) as a function of $V_g$ and $T$ (Figure 3d–f). By comparing $R_s$ contours under BGM and TGM, two features including lower $R_s$ and stronger $T$ dependence are generally shown under BGM. These remarkable discrepancies are a direct result of the lopsided cleanliness between the top and bottom dielectric interfaces.

In early studies,[13] we calculated the room-temperature $\mu$ for MoS$_2$ channels with different $t_s$ and found it is a unique strategy to modulate the charge distribution and $d$ values within channels. Also, we have pointed out in a review[3] the long-standing controversy on $\varepsilon_s$ of the 2D TMD channels, which represents the realistic screening capability to external Coulomb potential from the interfacial CIs. Recently, Kang et al[16] confirmed that $\varepsilon_s$ is non-fixed and highly dependent on $t_s$. Hence, there is likely great inaccuracy in previous theoretical calculations where a constant $\varepsilon_s$ was often adopted. In this context, it is highly necessary to perform a renewed experimental examination on the effects of Coulomb screening and scattering from interfacial CIs across the dimensional crossover.

To this end, we systematically collected the electronic transport data for the $t_s$-varied channels ranging from 1L to 5L under BGM and TGM with covering a broad $T$ range (8–300 K). Figure 4a,b shows the gating mode dependent $\mu$–$T$ curves for two typical thicknesses: a thin 1L and a thick 5L (see Figure S3 for the 2~4L samples). For both devices, $\mu$ exhibits band-like transport behavior, i.e., a negative $\mu$–$T$ relation, independent with gating modes, indicating the high channel quality. Upon reducing $T$, a general trend is $\mu$ first increases and then saturates below 40 K.

In case of the 1L MoS$_2$, $\mu$ depends hardly on gating modes at all $T$s. The values extracted under BGM and TGM are quite comparable, being ~30 and ~100 cm$^2$V$^{-1}$s$^{-1}$ at room and low $T$s, respectively, implying the weak overall screening effect to the interfacial CIs, owing to the extremely short $d_t$ and $d_b$ values (Figure 2a,b). Thus, the negligible change of $d$ from gating mode swapping is insufficient to cause a large variation in trends or magnitudes of $\mu$. In striking contrast, $\mu$ correlates closely with the gating modes in thick channels, especially in the low-$T$ regime. The value of the 5L sample changes significantly from ~190 ($\mu_{TGM}$) to ~500 cm$^2$V$^{-1}$s$^{-1}$ ($\mu_{BGM}$) at 8 K, featuring a 2.6-fold difference. In low-$T$ regime, CIs become the dominant scattering mechanism because the thermally excited lattice phonon mechanism is suppressed. Thus, in such a cleanliness lopsided structure, the dirtier TG dielectric interface, with crucial scattering parameters $n_t$ and $d_t$, acts as the leading scattering source in both gating modes. From Figure 2d,e, the scattering distances follow the relation $d_{t,BGM} > d_{t,TGM}$, which leads to the consequence $\mu_{BGM} > \mu_{TGM}$, because of the enhanced overall screening effect to CIs at larger $d_t$.

According to theoretical analyses,[3,12] among the leading scattering mechanisms limiting $\mu$ in high-quality 2D TMD channels, lattice phonons and interfacial CIs are mainly responsible for high- and low-$T$ regimes, respectively. In the high-$T$ regime, typically above 100 K, $\mu$ follows a $\mu \propto T^{-\gamma}$ power law[21] and the magnitude of $\gamma$ was often used to analyze the contribution of CIs. Theoretical studies predict that, in the case of pure phonon scattering, $\gamma$ ~ 1.7 and ~2.6 for 1L and bulk MoS$_2$, respectively,[22] and it would decrease and deviate obviously from the ideal values if the CI contribution is strong. For the 1L sample (Figure S4a), we extract reduced $\gamma$ values of 0.75–0.85 for both gating modes, much lower than the ideal pure-phonon-dominated value.[14,22] This indicates CI scattering also plays an important role at the high-$T$ region and cannot be suppressed by varying gating modes (i.e., via modulating $d$) for the 1L channels. In contrast, $\gamma$ can be enhanced from 0.8 (TGM) to 1.5 (BGM) for the 5L counterpart, being pushed close to the theoretical prediction. Similar trends are also observed for the 3–4L MoS$_2$. This observation has two implications: 1) the TG dielectric interface is much dirtier than that of BG, that is $n_t \gg n_b$, resulting likely from the partially oxidized seeding layers and chemical residues introduced during processing; 2) The carriers induced near the BG dielectric undergo a largely reduced potential from top dielectric surface after the natural screening by the 5L channel body (see Figure 2d). Through theoretical fittings, we estimate a low $n_b \sim 10^{12}$ and a high $n_t \sim 10^{13}$ cm$^{-2}$ at the bottom and top dielectric interfaces, respectively, showing a 10-fold difference in interface cleanliness.

To gain more insight into the effects of lopsided dielectric cleanliness and $d$ variation on Coulomb scattering, we calculate and compare the theoretical $\mu - T$ trends for individual scattering components. According to the Boltzmann transport theory and random phase approximation, the CI scattering rate can be written as

$$\frac{1}{\tau_m(\mathbf{k})} = \frac{2\pi}{\hbar} n_i \sum_{\mathbf{q}} V_q^2 (1-\cos\theta) \delta(E_{\mathbf{k}} - E_{\mathbf{k}'}), \tag{1}$$

where $n_i$ is the CI concentration (either $n_b$ or $n_t$ in this work), $q = 2k\sin(\theta/2)$ is the scattering vector with $\theta$ being the scattering angle from the initial momentum $\mathbf{k}$ to the final momentum $\mathbf{k}'$. In Equation 1, the scattering matrix element $V_q$ is strongly associated with the confutative form factors $F_{CI}(\varepsilon_s, d)$ and polarization function $\varepsilon_{2D}(\varepsilon_s, d)$, which can be expressed as

$$V_q = \frac{e^2 F_{CI}(q)}{2\varepsilon_s \varepsilon_{2D}(q) q}. \tag{2}$$

In contrast to most works that employ simplified treatments to derive $F(\varepsilon_s, d)$ and $\varepsilon_{2D}(\varepsilon_s, d)$ by using the delta or trigonometric approximation to mimic carrier distributions and adopting a fixed $\varepsilon_s$ value for all thicknesses, we employed the realistic distribution functions and $\varepsilon_s$ values for each $t_s$, and performed strict numerical calculations to treat these two critical Coulomb terms, which enables us to precisely capture the subtle parametric changes in $\varepsilon_s$ and $t_s$ across dimensional crossover and the resultant impacts on Coulomb screening and scattering. See Equations A1–A10 in Supporting Information for more details on the theoretical treatments.

Figure 4c–f shows the theoretical curves from the five primary scattering components, including polar longitudinal optical phonons (PLO), acoustic deformation potential (ADP), surface optical phonons (SO), short-range lattice defects (SR), and CI at top and bottom dielectrics (CI_top, CI_bot). We note that both the magnitudes and trends of $\mu$ from PLO, ADP, SO and SR are irrelevant to gating modes and, thus, the corresponding $\mu$–$T$ curves are identical under different gating modes and only the two CI-relevant $\mu$–$T$ curves change, due to the modulation on $n_{CI}$ and $d$ upon swapping the gating mode.

We then analyze the Coulomb scattering by comparing the CI relevant $\mu$–$T$ curves (red and blue lines in Figure 4c–f) between different gating modes or among varied channels, where the primary $n_{CI}$ and $d$ parameters are modified. Several features can be seen. First, the two CI-related curves show distinct responses in the 1L and 5L devices when swapping the gating modes between BGM and TGM. For the 1L device, the CI_top and CI_bot curves exhibit inconspicuous changes (Figure 4c,e), due to the negligible change in $d$ as shown in Figure 2a,b. In contrast, they greatly reshape themselves in the 5L devices, showing a large variation in magnitude (Figure 4d,f), because of the synergic modifications on $d_t$ and $d_b$ and the large difference between $n_t$ and $n_b$ (Figure 2d,e). Second, under the same gating and $T$ conditions, the Coulomb scattering in the 1L device is generally stronger than that in the 5L device, as manifested by the reduced $\mu$ magnitudes. This observation is associated with the reduced $d_t$ and $d_b$ in the thinner channels due to the concentrated charge distribution, which can be understood by comparing Figure 2a with Figure 2d.

Finally, we remark on the dependence of static $\varepsilon_s$ on $t_s$ and its impact on $\mu$. According to previous reports,[16–18] $\varepsilon_s$ can decrease from 12 to 4.8 as MoS$_2$ is reduced from bulk to 1L. To data, however, no works have analyzed such a dimensional crossover effect on the actual dielectric screening in TMDs. Here, we calculate the low-$T$ $\mu_{BGM}$ and $\mu_{TGM}$ as a function of $t_s$ for three types of $\varepsilon_s$ values, that is, by employing two constant values, 4.8 and 12.8, which are adopted from 1L and bulk and represents the lower and upper limits, for all channels, and the realistic $t_s$-dependent values.[19] Figure 5a,b show measured and calculated $\mu_{TGM}$ and $\mu_{BGM}$ against $t_s$ at 10 K by using the above three types of $\varepsilon_s$ values. A counterintuitive finding is that the overall screening effect, including dielectric screening and electron-electron interaction, is stronger, rather than weaker, in channels with lower $\varepsilon_s$, because the magnitudes of $\mu_{BGM}$ and $\mu_{TGM}$ derived with $\varepsilon_s=4.8$ are both generally higher than those with $\varepsilon_s=12.8$ for all adopted $t_s$s, implying a quicker variation in the form factor $F(\varepsilon_s, d)$ than the dielectric function $\varepsilon_{2D}(\varepsilon_s, d)$, in response to the $\varepsilon_s$ variation. For the 1L device, the discrepancies in $\mu$ are 5% and 10% under BGM and TGM, respectively, while they increase to 30% and 40% for the 5L device when constant $\varepsilon_s$ values are used in the calculation. Thus, it is important to adopt appropriate $\varepsilon_s$ values according to $t_s$. As expected, the calculations agree best with the measured $\mu - t_s$ data when variable $\varepsilon_s$ values are used with globally assuming $n_t = 10^{13}$ and $n_b = 10^{12}$ cm$^{-2}$ (Figure 5a,b). At last, by adopting the appropriate $\varepsilon_s$ values for each channel across the crossover region from 1L to 5L, we calculated the $\mu$–$T$ curves under different gating modes (Figure 5c,d), where the overall $\mu$ values are obtained by summing up all the five scattering components with the Matthiessen's rule. A notable feature is the distinct low-$T$ $\mu$ behavior with increasing $t_s$, which exhibits a continuous rise under BGM but a saturation under TGM till 5L. Such contrasting trends result from the swap between the main CI sources located at the cleanliness-asymmetric dielectric interfaces, which can be easily understood from the schematic carrier distributions plotted in Figure 2.

We performed jointly experimental and theoretical studies on the realistic effects of Coulomb screening and scattering in atomically thin MoS$_2$ channels covering the 2D and 3D regimes. A unique dual-gated FET architecture with asymmetric dielectric cleanliness was employed to facilitate the controls on the density of leading Coulomb scattering sources by swapping gating modes. It was found that the redistribution of charges within channels controlled by gating modes can result in remarkable changes in the density of Coulomb impurities and overall scattering intensity and, thus, electronic performance. Together with the variation of channel thickness, another two crucial screening factors, including interaction distance and electric permittivity were also flexible modulated. In the analysis of the trends of charge mobility versus temperature and channel thickness, we unfolded the specific effects of screening and scattering in TMD channels. The contrastive analyses on the electronic performance, via multiparametric modulations on scattering factors, provide an unprecedentedly insightful understanding of the Coulomb screening and scattering effects in atomically thin semiconductors.

**Supporting Information**

Additional details on device fabrication, capacitance measurement, Raman and electronic characterization, and theoretical expressions for various scattering mechanisms.

**Acknowledgements**

This work was supported by the National Natural Science Foundation of China (61974060 and 61674080), the National Key R&D Program of China (2021YFA1202903), the Innovation and Entrepreneurship Program of Jiangsu province, and the Micro Fabrication and Integration Technology Center in Nanjing University.

**References**

1. Liu, Y.; Duan, X.; Shin, H.-J.; Park, S.; Huang, Y.; Duan, X. Promises and Prospects of Two-Dimensional Transistors. *Nature* **2021**, *591*, 43–53.

2. Li, M.-Y.; Su, S.-K.; Wong, H.-S. P.; Li, L.-J. How 2D Semiconductors Could Extend Moore's Law. *Nature* **2019**, *567*, 169–170.

3. Li, S.-L.; Tsukagoshi, K.; Orgiu, E.; Samorì, P. Charge Transport and Mobility Engineering in Two-Dimensional Transition Metal Chalcogenide Semiconductors. *Chem. Soc. Rev.* **2016**, *45*, 118–151.

4. Cai, X.; Wu, Z.; Han, X.; Chen, Y.; Xu, S.; Lin, J.; Han, T.; He, P.; Feng, X.; An, L.; Shi, R.; Wang, J.; Ying, Z.; Cai, Y.; Hua, M.; Liu, J.; Pan, D.; Cheng, C.; Wang, N. Bridging the Gap between Atomically Thin Semiconductors and Metal Leads. *Nat. Commun.* **2022**, *13*, 1777.

5. Cui, X.; Lee, G.-H.; Kim, Y. D.; Arefe, G.; Huang, P. Y.; Lee, C.-H.; Chenet, D. A.; Zhang, X.; Wang, L.; Ye, F.; Pizzocchero, F.; Jessen, B. S.; Watanabe, K.; Taniguchi, T.; Muller, D. A.; Low, T.; Kim, P.; Hone, J. Multi-Terminal Transport Measurements of $MoS_2$ Using a van der Waals Heterostructure Device Platform. *Nat. Nanotechnol.* **2015**, *10*, 534–540.

6. Wang, Y.; Kim, J. C.; Wu, R. J.; Martinez, J.; Song, X.; Yang, J.; Zhao, F.; Mkhoyan, K. A.; Jeong, H. Y.; Chhowalla, M. Van der Waals Contacts between Three-Dimensional Metals and Two-Dimensional Semiconductors. *Nature* **2019**, *568*, 70–74.

7. Liu, Y.; Guo, J.; Zhu, E.; Liao, L.; Lee, S.-J.; Ding, M.; Shakir, I.; Gambin, V.; Huang, Y.; Duan, X. Approaching the Schottky-Mott Limit in van der Waals Metal-Semiconductor Junctions. *Nature* **2018**, *557*, 696–700.

8. Shen, P.-C.; Su, C.; Lin, Y.; Chou, A.-S.; Cheng, C.-C.; Park, J.-H.; Chiu, M.-H.; Lu, A.-Y.; Tang, H.-L.; Tavakoli, M. M.; Pitner, G.; Ji, X.; Cai, Z.; Mao, N.; Wang, J.; Tung, V.; Li, J.; Bokor, J.; Zettl, A.; Wu, C.-I.; Palacios, T.; Li, L.-J.; Kong, J. Ultralow Contact Resistance between Semimetal and Monolayer Semiconductors. *Nature* **2021**, *593*, 211–217.

9. Desai, S. B.; Madhvapathy, S. R.; Sachid, A. B.; Llinas, J. P.; Wang, Q.; Ahn, G. H.; Pitner, G.; Kim, M. J.; Bokor, J.; Hu, C.; Wong, H.-S. P.; Javey, A. $MoS_2$ Transistors with 1-Nanometer Gate Lengths. *Science* **2016**, *354*, 99–102.

10. Wu, F.; Tian, H.; Shen, Y.; Hou, Z.; Ren, J.; Gou, G.; Sun, Y.; Yang, Y.; Ren, T.-L. Vertical $MoS_2$ Transistors with Sub-1-nm Gate Lengths. *Nature* **2022**, *603*, 259–264.

11. Li, S.-L.; Tsukagoshi, K. Carrier Injection and Scattering in Atomically Thin Chalcogenides. *J. Phys. Soc. Jpn.* **2015**, *84*, 121011.

12. Yu, Z.; Ong, Z.-Y.; Li, S.; Xu, J.-B.; Zhang, G.; Zhang, Y.-W.; Shi, Y.; Wang, X. Analyzing the Carrier Mobility in Transition-Metal Dichalcogenide $MoS_2$ Field-Effect Transistors. *Adv. Funct. Mater.* **2017**, *27*, 1604093.

13. Li, S.-L.; Wakabayashi, K.; Xu, Y.; Nakaharai, S.; Komatsu, K.; Li, W.-W.; Lin, Y.-F.; Aparecido-Ferreira, A.; Tsukagoshi, K. Thickness-Dependent Interfacial Coulomb Scattering in Atomically Thin Field-Effect Transistors. *Nano Lett.* **2013**, *13*, 3546–3552.

14. Ma, N.; Jena, D. Charge Scattering and Mobility in Atomically Thin Semiconductors. *Phys. Rev. X* **2014**, *4*, 011043.

15. Ando, T.; Fowler, A. B.; Stern, F. Electronic Properties of Two-Dimensional Systems. *Rev. Mod. Phys.* **1982**, *54*, 437–672.

16. Kang, Y.; Jeon, D.; Kim, T. Local Mapping of the Thickness-Dependent Dielectric Constant of $MoS_2$. *J. Phys. Chem. C* **2021**, *125*, 3611–3615.

17. Yim, C.; OBrien, M.; McEvoy, N.; Winters, S.; Mirza, I.; Lunney, J. G.; Duesberg, G. S. Investigation of the Optical Properties of $MoS_2$ Thin Films Using Spectroscopic Ellipsometry. *Appl. Phys. Lett.* **2014**, *104*, 103114.

18. Chen, X.; Wu, Z.; Xu, S.; Wang, L.; Huang, R.; Han, Y.; Ye, W.; Xiong, W.; Han, T.; Long, G.; Wang, Y.; He, Y.; Cai, Y.; Sheng, P.; Wang, N. Probing the Electron States and Metal-Insulator Transition Mechanisms in Molybdenum Disulphide Vertical Heterostructures. *Nat. Commun.* **2015**, *6*, 6088.

19. Kumar, A.; Ahluwalia, P. Tunable Dielectric Response of Transition Metals Dichalcogenides $MX_2$ (M=Mo, W; X=S, Se, Te): Effect of Quantum Confinement. *Physica B* **2012**, *407*, 4627–4634.

20. Fuhrer, M. S.; Hone, J. Measurement of Mobility in Dual-Gated $MoS_2$ Transistors. *Nat. Nanotechnol.* **2013**, *8*, 146–147.

21. Baugher, B. W. H.; Churchill, H. O. H.; Yang, Y.; Jarillo-Herrero, P. Intrinsic Electronic Transport Properties of High-Quality Monolayer and Bilayer $MoS_2$. *Nano Lett.* **2013**, *13*, 4212–4216.

22. Kaasbjerg, K.; Thygesen, K. S.; Jacobsen, K. W. Phonon-Limited Mobility in n-Type Single-Layer $MoS_2$ from First Principles. *Phys. Rev. B* **2012**, *85*, 115317.

**Figure captions**

Figure 1. (a–d) Schematic fabrication process for dual-gated $MoS_2$ FETs. $V^+$ and $V^-$ denote the two inner voltage probes used in the four-terminal measurement. D: drain, S: source, BG: bottom gate, TG: top gate. (e, f) Optical images for a 2L $MoS_2$ FET before and after deposition of TG stacks. The dotted red and blue lines denote the areas of channel and TG stacks, respectively. Scale bar: 10 μm. (g) Schematic 3D view for dual-gated $MoS_2$ FETs. (h–j) Surface morphologies taken by AFM images for bottom $Al_2O_3$, transferred $MoS_2$, and seeding layer before ALD, respectively. Scale bar: 5 μm. (k) Typical transfer curves of a 2L $MoS_2$ channel under three different gating modes including DGM, BGM, and DGM.

Figure 2. (a–c) Schematic diagrams for wiring geometry, carrier distribution, and interaction distances under different gating modes for the dual-gated 1L $MoS_2$ FET. $t$: channel thickness; CI_top (bot): Coulomb impurity at top (bottom) dielectric interfaces; $d_t$, $d_b$: scattering distance to the top and bottom CIs, respectively. (d–f) Corresponding diagrams for a FET with a thick 5L $MoS_2$ channel.

Figure 3. (a–c) $T$-dependent four-terminal $\sigma$ versus $V_g$ for a typical 5L $MoS_2$ channel under different gating modes BGM, TGM, and DGM, respectively. For comparison, the data in (a) and (b) are magnified by 1.5 and 4 times, respectively. (d–f) Contour plots for $R_s$ as a function of $V_g$ and $T$ for the 5L $MoS_2$ channel under different gating modes.

Figure 4. (a, b) Comparison of experimental (dots) and theoretical (lines) $\mu$-$T$ curves for the 1L and 5L $MoS_2$ FETs. All $\mu$ values are adopted at $n_{2D} = 10^{13}$ cm$^{-2}$. In theoretical calculation, the CI densities $n_t$ and $n_b$ are assumed to be $10^{13}$ and $10^{12}$ cm$^{-2}$, respectively. (c–f) Theoretically calculated $\mu$-$T$ curves for the five leading scattering mechanisms for the 1L and 5L $MoS_2$ channels and under different gating modes. CI_top and CI_bot: CIs located at the top and bottom dielectrics; ADP: acoustic deformation potential; PLO: polar longitudinal optical phonons; SR: short-range lattice defects; SO: surface optical phonons.

Figure 5. (a, b) Comparison of experimental low-$T$ $\mu$ data (blue and red symbols) with theoretical calculations (lines) under different gating modes, based on assumed constant and practically $t_s$-dependent $\varepsilon_s$ values across the dimensional crossover from 1L to 5L. All $\mu$ values are adopted at $n_{2D} = 10^{13}$ cm$^{-2}$. $n_t = 10^{13}$ and $n_b = 10^{12}$ cm$^{-2}$ are used in calculation. (c, d) Theoretically calculated overall $\mu$-$T$ curves for the channels from 1L to 5L under different gating modes.

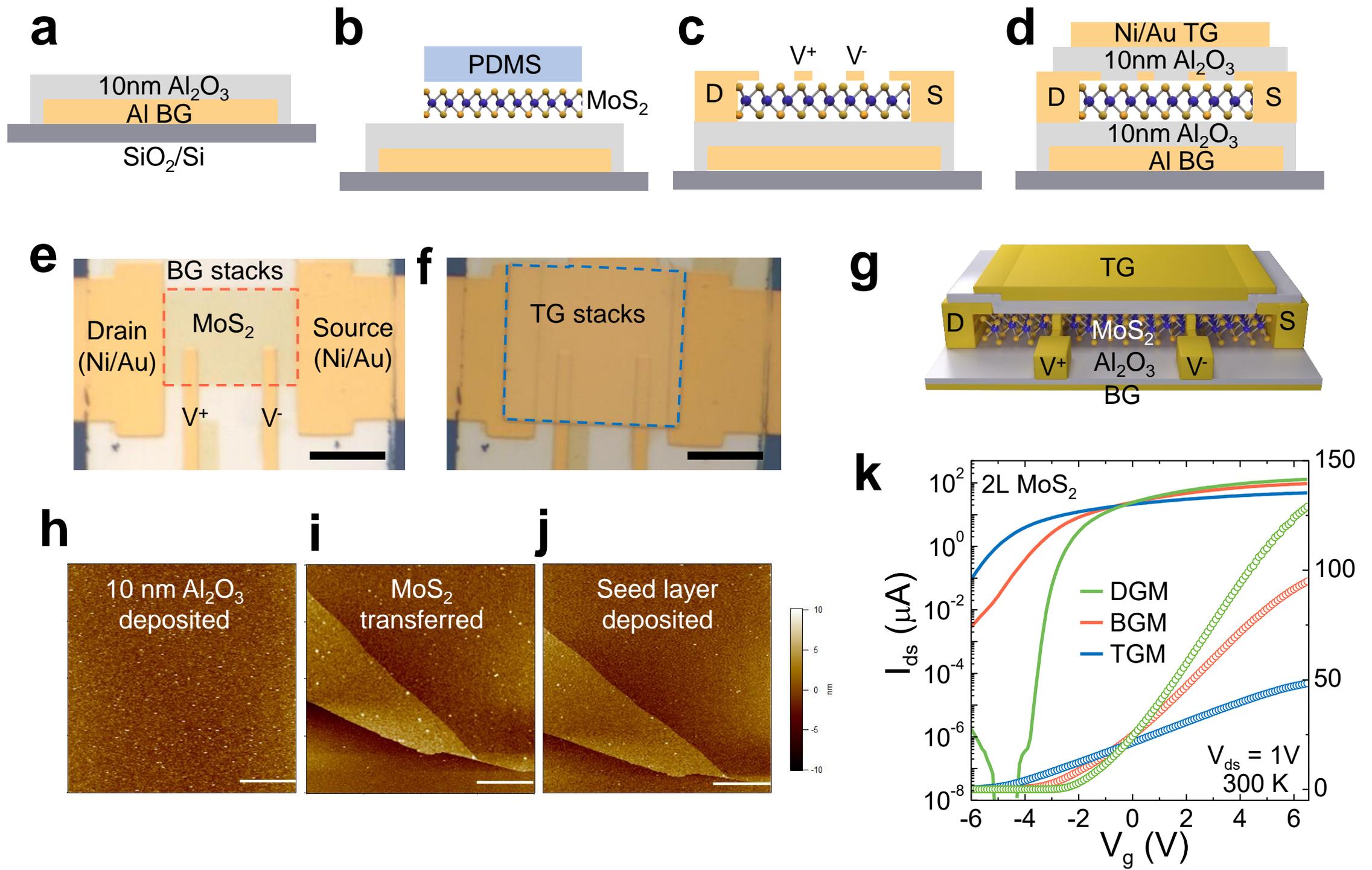

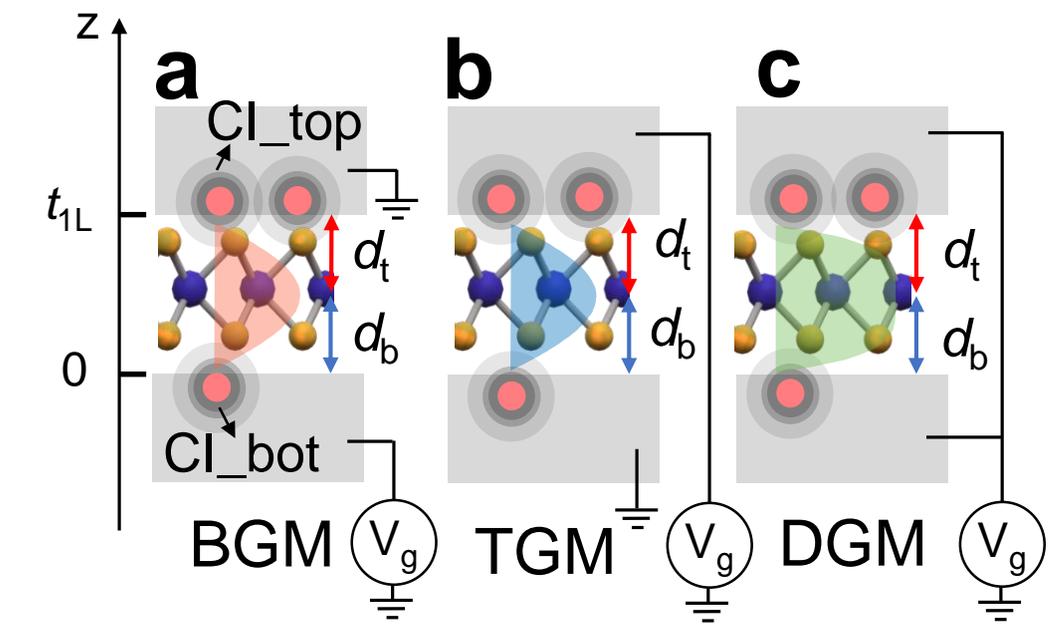
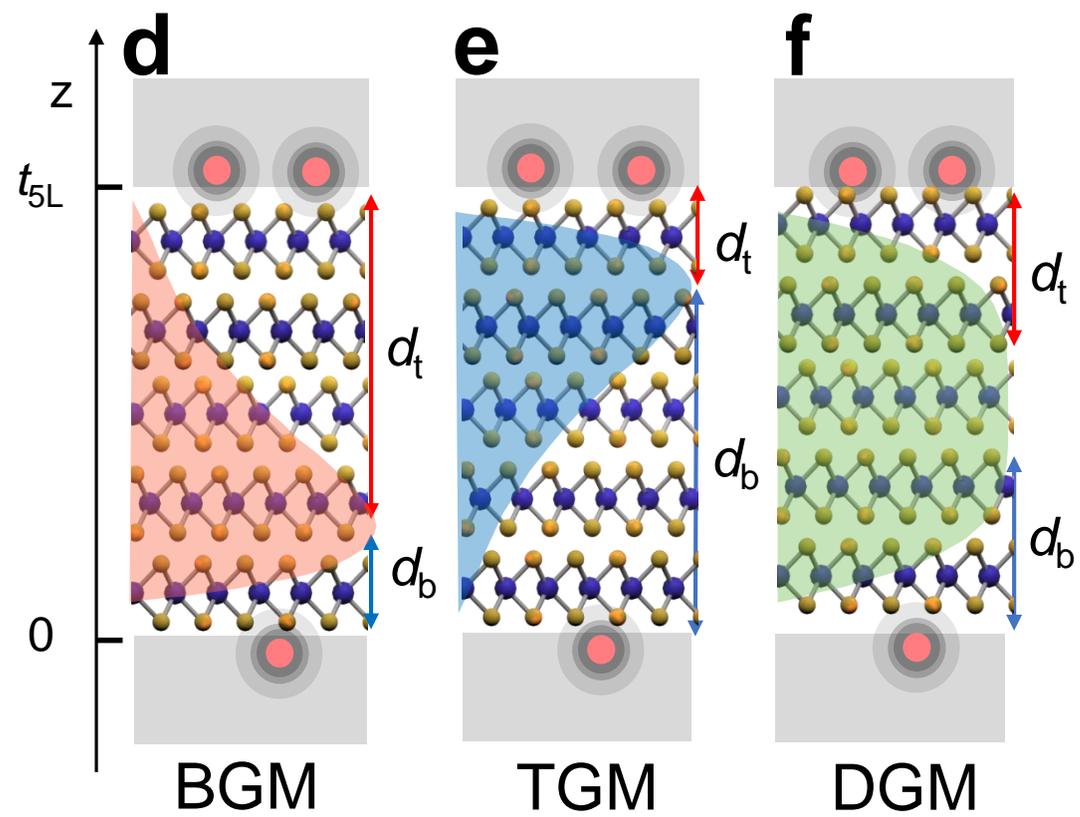

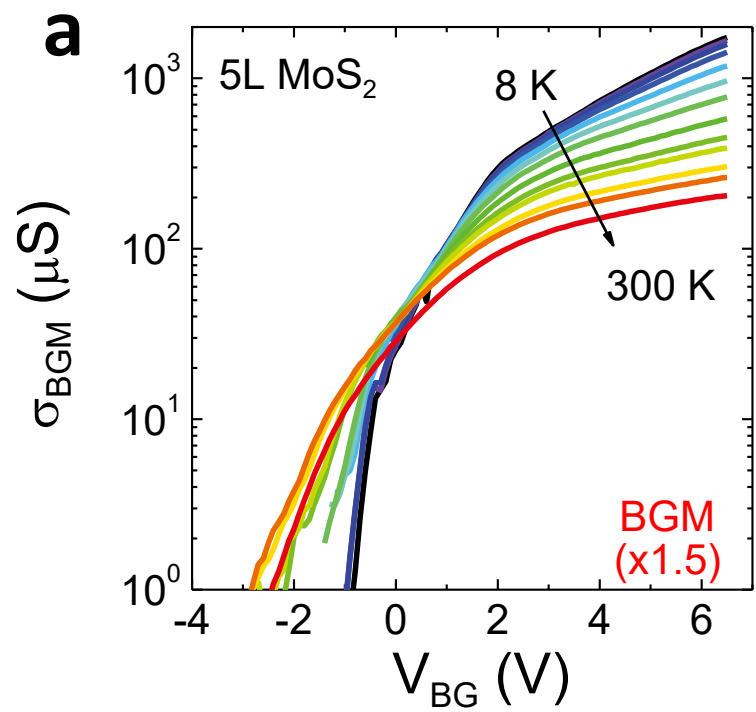
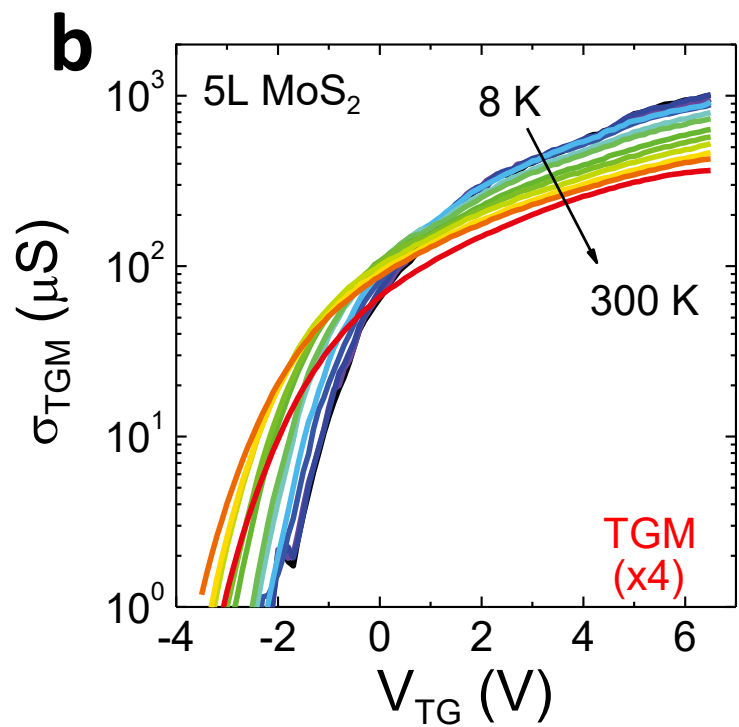
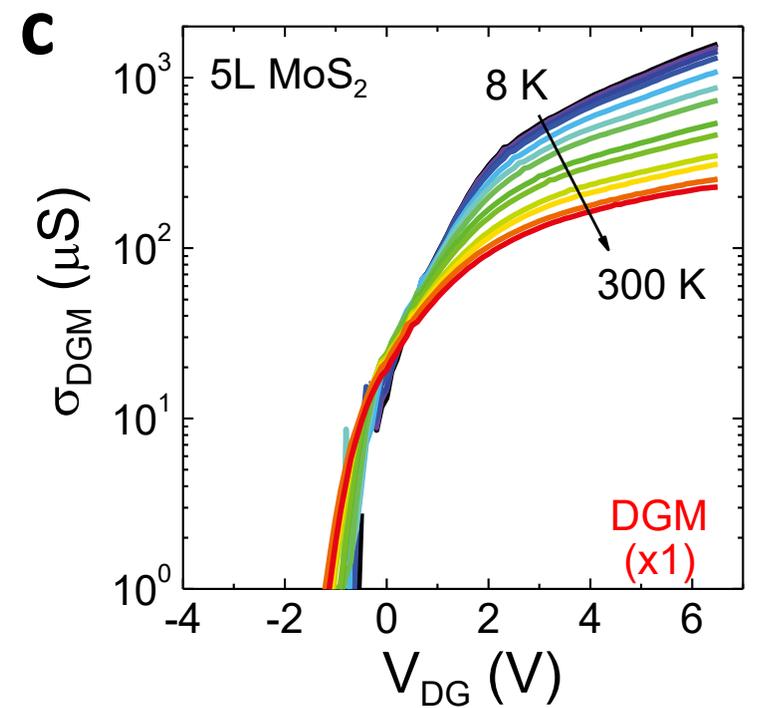
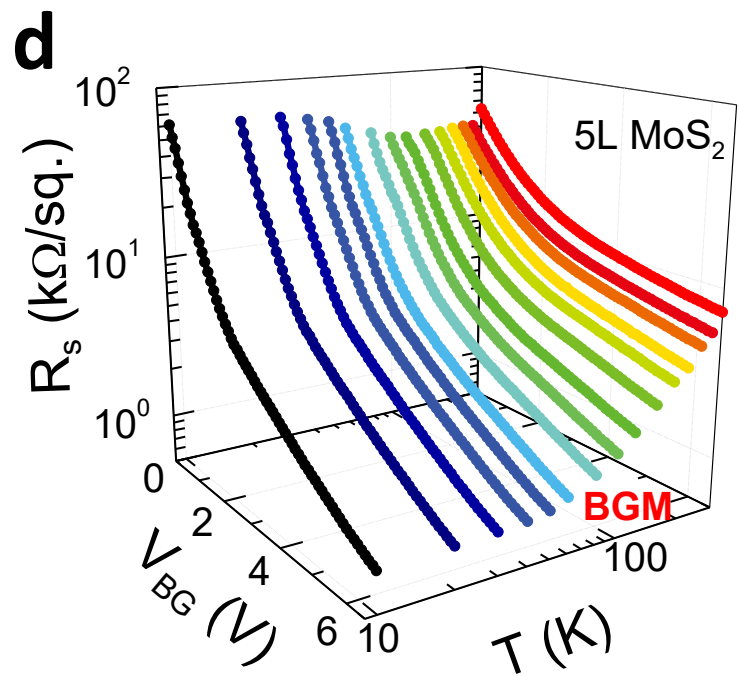
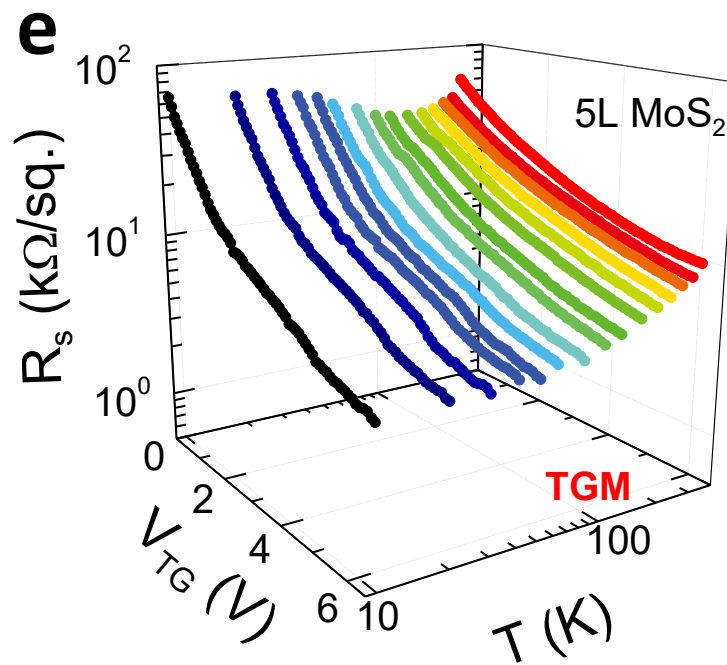
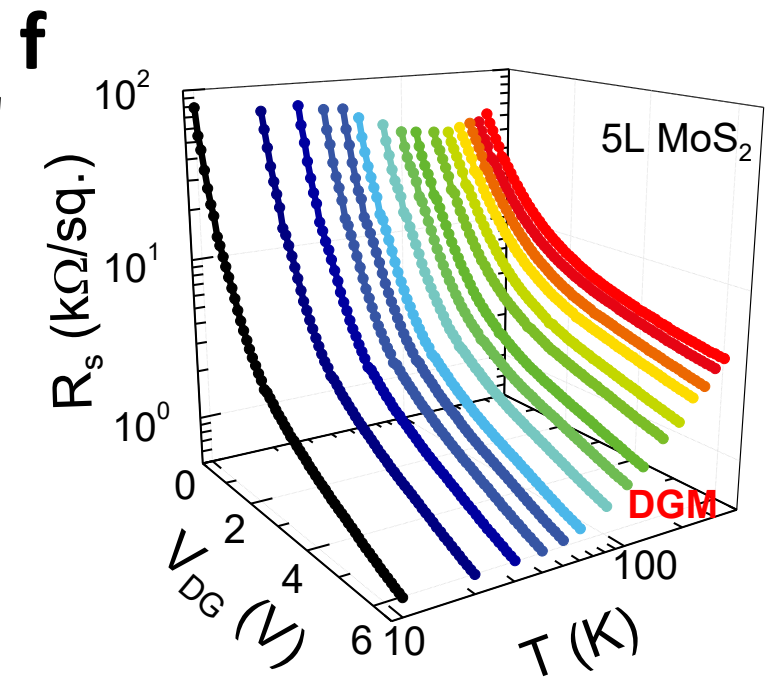

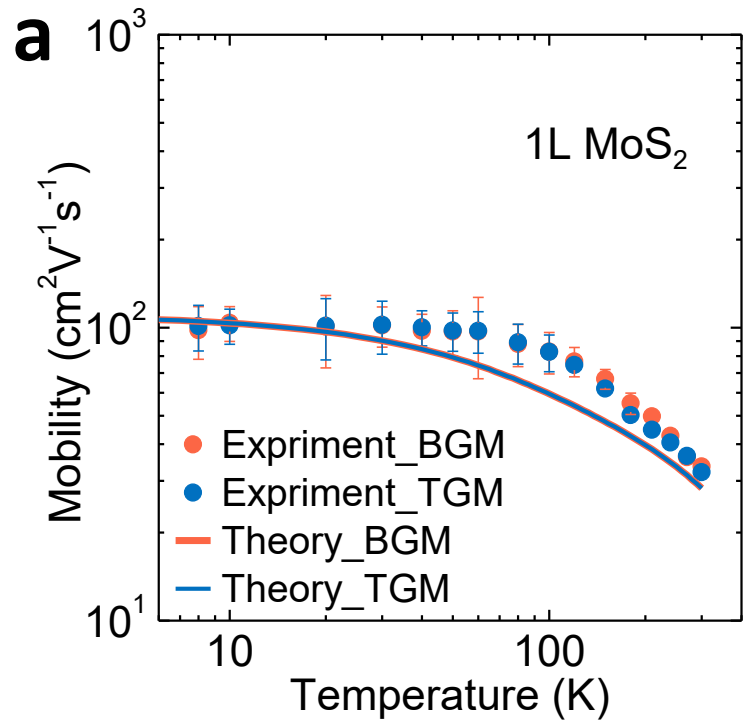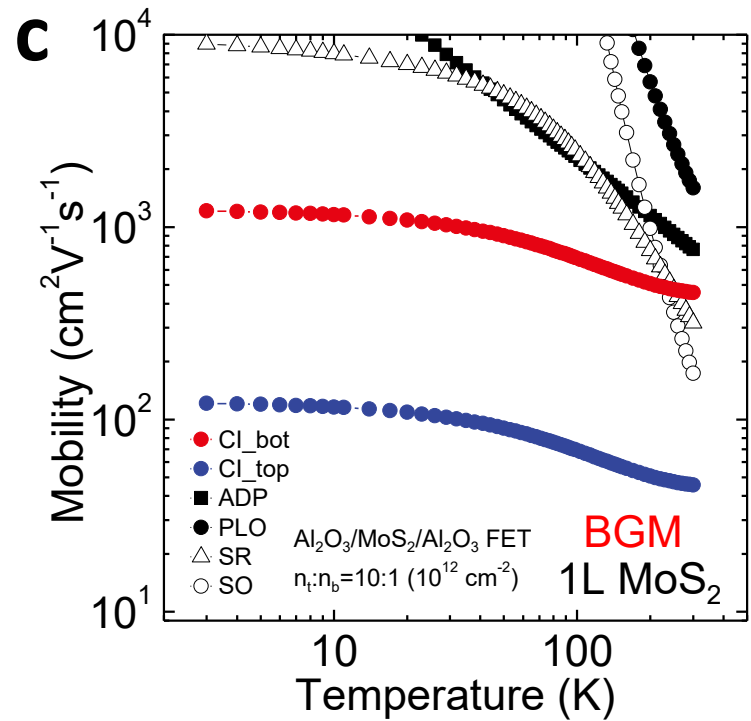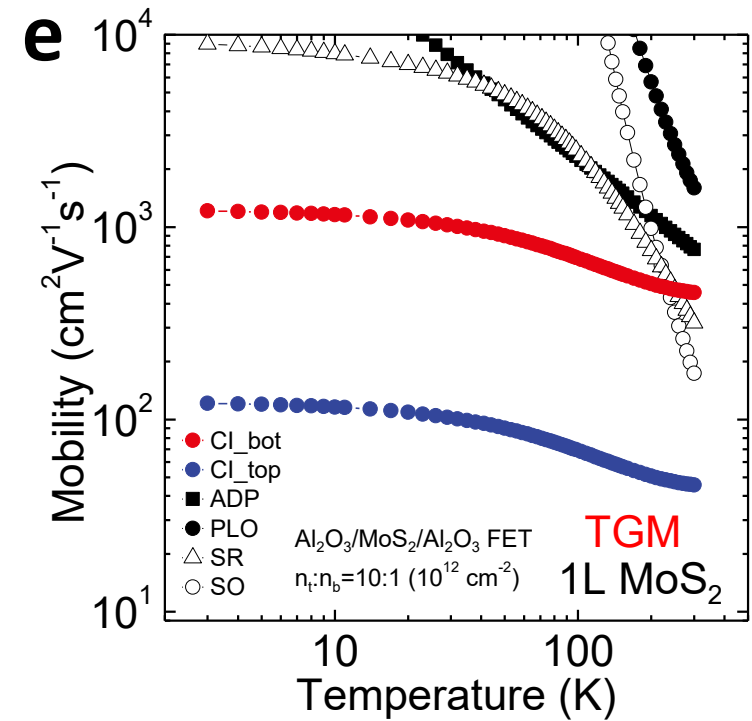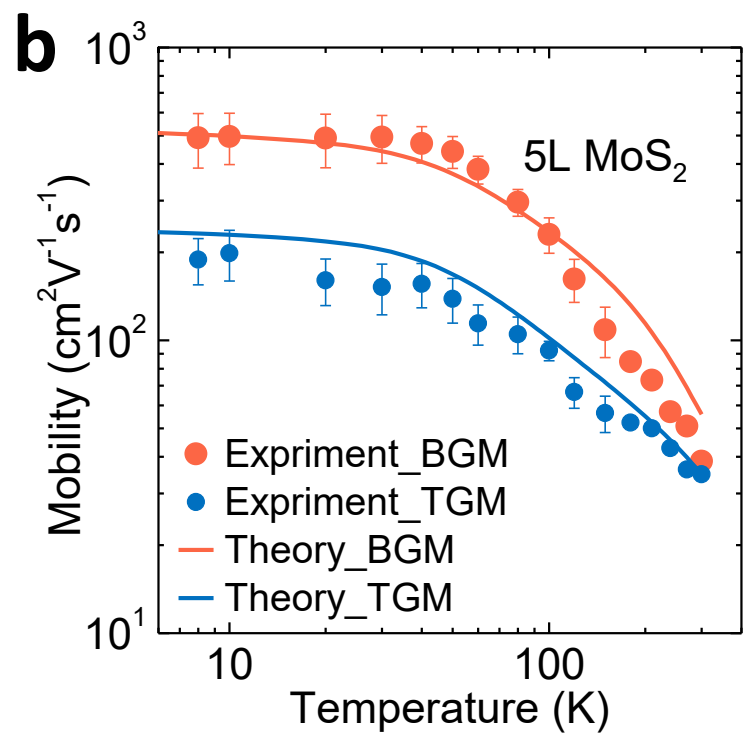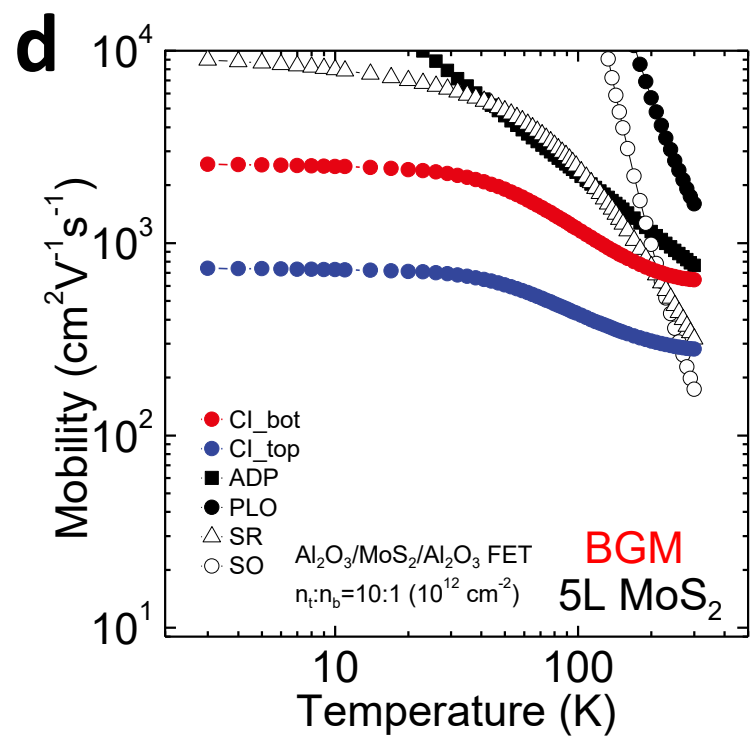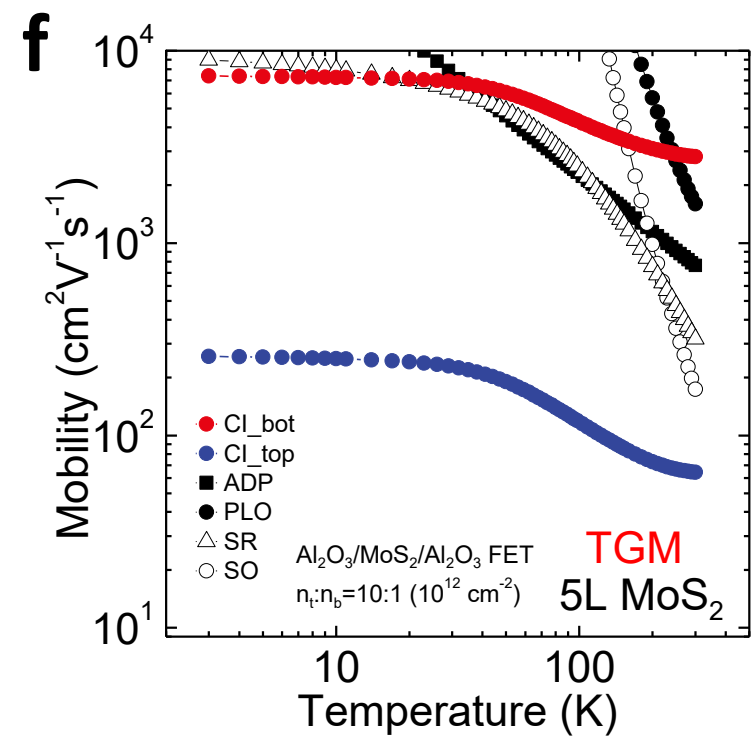

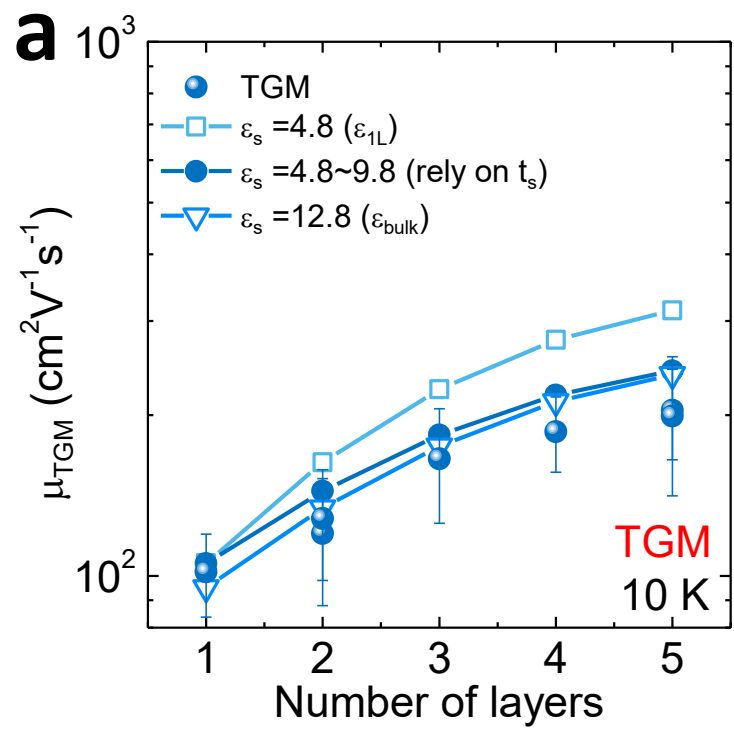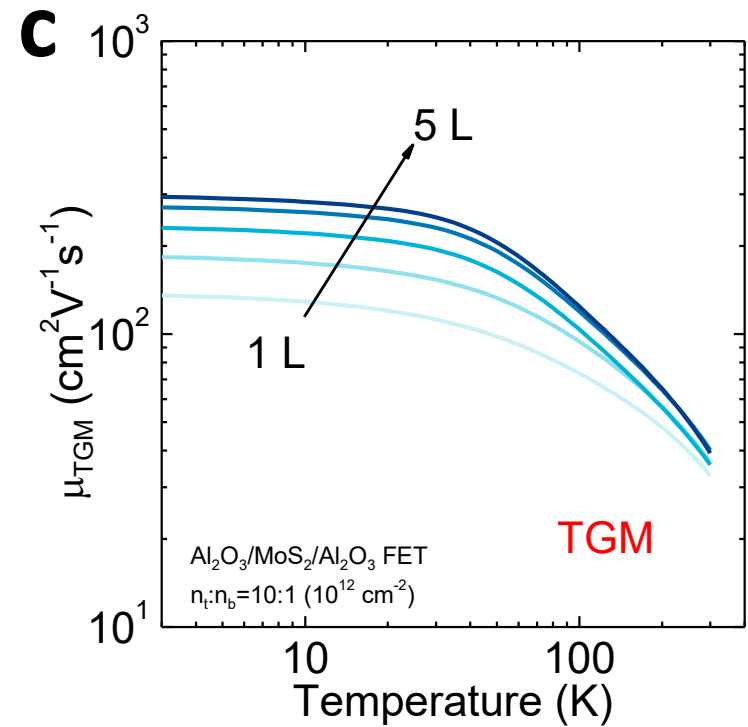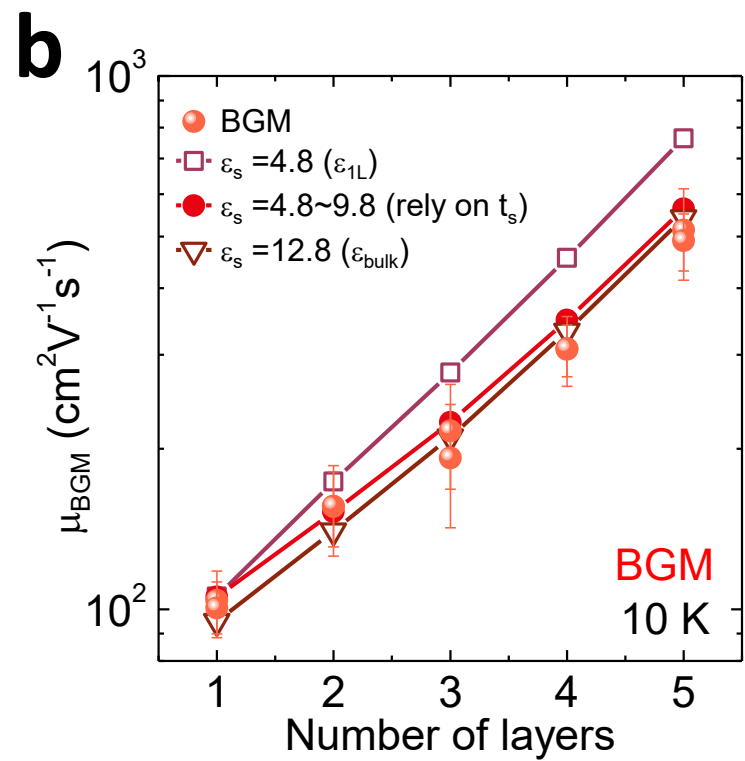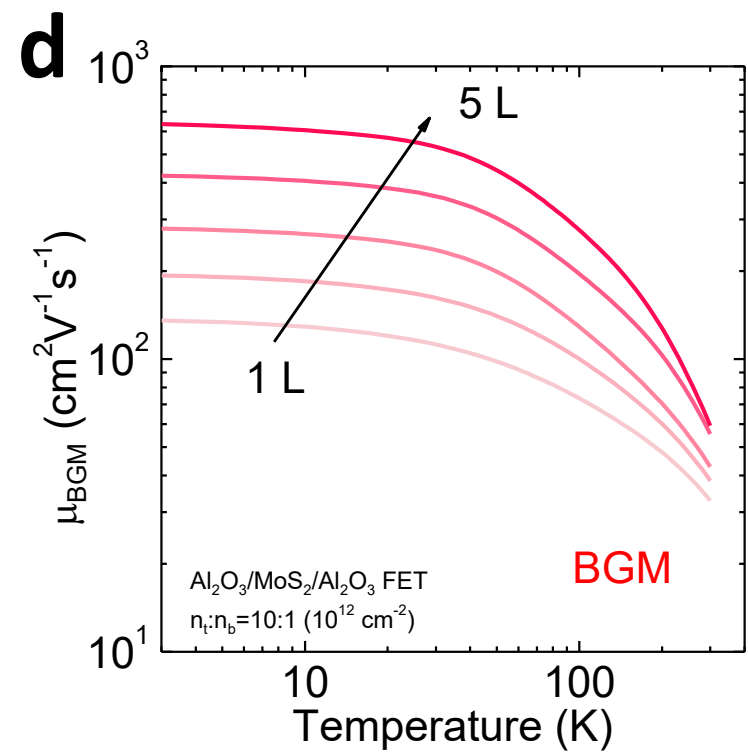

# Coulomb screening and scattering in atomically thin transistors across dimensional crossover

Shihao Ju, Binxi Liang, Jian Zhou, Danfeng Pan, Yi Shi, Songlin Li*

*National Laboratory of Solid-State Microstructures, Collaborative Innovation Center of Advanced Microstructures, and School of Electronic Science and Engineering, Nanjing University, Nanjing, Jiangsu 210023, China*

*Author to whom correspondence should be addressed: sli@nju.edu.cn

## 1. Fabrication of dual-gated MoS$_2$ FETs

At first, a layer of 10 nm Al$_2$O$_3$ was deposited as BG dielectrics onto Al strips by atomic layer deposition (ALD). The ALD process was performed at 150 °C using tri(methyl) aluminum (TMA) and H$_2$O as precursors, with high purity N$_2$ as carrier gas at a flow rate of 20 sccm. The pulse times of TMA and H$_2$O were both set at 30 ms and the purge times for both precursors were 30 s. The resultant growth rate of Al$_2$O$_3$ was ~1 Å/cycle. The MoS$_2$ flakes were mechanically exfoliated by blue tapes from natural molybdenite minerals. Viscous organic polymer polydimethylsiloxane (PDMS) was utilized to further exfoliate the MoS$_2$ flakes and dry transfer them from the blue tapes onto the local BG stacks. Afterward, Ni/Au (5/60 nm) electrodes were defined on MoS$_2$ flakes as electrodes by standard electron beam lithography and thermal evaporation. Next, 1 nm Al was deposited at ~0.5 Å/s on MoS$_2$ channels as the seeding layer for growing TG dielectrics on dangling-bond-free MoS$_2$ surfaces. Finally, a 10 nm Al$_2$O$_3$ layer and 5/60 nm Ni/Au layer were deposited in sequence by ALD and thermal evaporation to form the TG stacks.

## 2. Electrical and thickness characterization

All electrical characterization was performed with Keithley 2636B Sourcemeters under a high vacuum (pressure < 10$^{-5}$ Torr) with a cryogenic probe station (CRX-6.5K, Lakeshore). To check the role of the idle gate and the channel depletion characteristic in the single gating modes (i.e., BGM and TGM), we measured two types of transfer curves with the idle gate grounded or floated when sweeping the operating gate (Figure S1a,b). Evidently, the 2L MoS$_2$ channel can be fully switched OFF (i.e., fully depleted) when the idle gate is floated (~ applied $V_g$), while it remains partially undepleted when the idle gate is grounded (0 V). Therefore, the incomplete OFF states in Figure 1k result from the grounding mode of the idle gate in the single gating modes (BGM and TGM), rather than the low gating efficiency. Also, the gate leakage of the 10-nm Al$_2$O$_3$ dielectrics was checked (Figure S1c). The 10-nm Al$_2$O$_3$ dielectric can endure $V_g$ from -6 to +6V, with the gate leakage below 5 pA at -6V and less than 25 pA at +6.5 V.

The thickness information of the MoS$_2$ samples was identified by the Raman spectrum (Figure S1d), according to the distance between the $E_{2g}$ and $A_{1g}$ modes or the ratios of them to the Si mode at 520 cm$^{-1}$.[1]

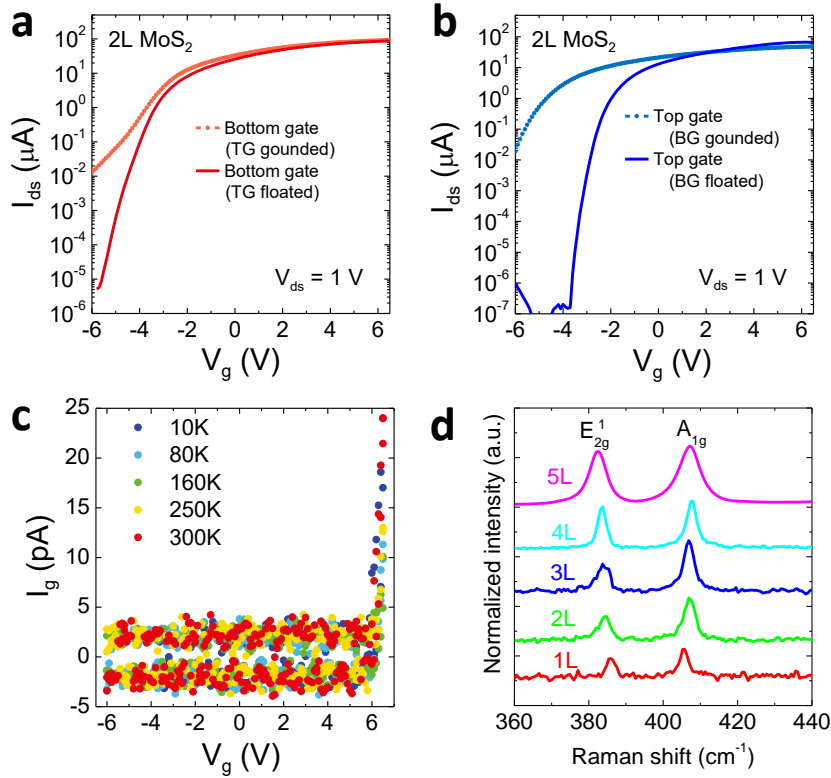

Figure S1. (a, b) Channel depletion characteristics in the single gating modes (BGM and TGM) verified from the contrastive transfer curves collected with the idle gate grounded or floated when sweeping the operating gate. (c) Gate leakage at different biasing and temperature conditions. (d) Raman spectra for the 1L–5L MoS$_2$ on Al$_2$O$_3$ dielectrics.

## 3. Capacitance measurements for bottom and top dielectrics

Figure S2a–c shows the wiring geometry used in capacitance measurements under the three gating modes (i.e., BGM, TGM, and DGM). Figure S2d shows the capacitance-voltage (C-V) characteristics at the frequency of 1kHz. An increase in capacitance is seen when applying positive gate voltages, due to the gating-induced carrier accumulation in channels. The absolute capacitances recorded for the BG and TG dielectrics are 11.4 pF and 3.2 pF, respectively (Figure S2e). Given the overlaps between the BG and source/drain electrodes, which bring about parasitic capacitances, the actual gate areas for BG and TG are ~1780 and ~710 μm$^2$ (Figure S2d). The estimated coupling capacitances per unit area for $C_{BG}$ and $C_{TG}$ are 0.64 and 0.45 μF·cm$^{-2}$, respectively. The slightly low value of $C_{TG}$ can be ascribed to the oxidation of the seeding layer, which forms an additional dielectric layer rather than ALD. In Figure S2e, the as-recorded capacitive values under DGM are also consistent with the summation of these under TGM and BGM. To verify the ratio between TG and BG dielectrics, we further draw the contour plot of the channel conductance versus $V_{BG}$ and $V_{TG}$. According to the isograms, a slope around 0.72 is also extracted, which agrees with $C_{TG}/C_{BG}$ ratio (0.45/0.64) from measurements.

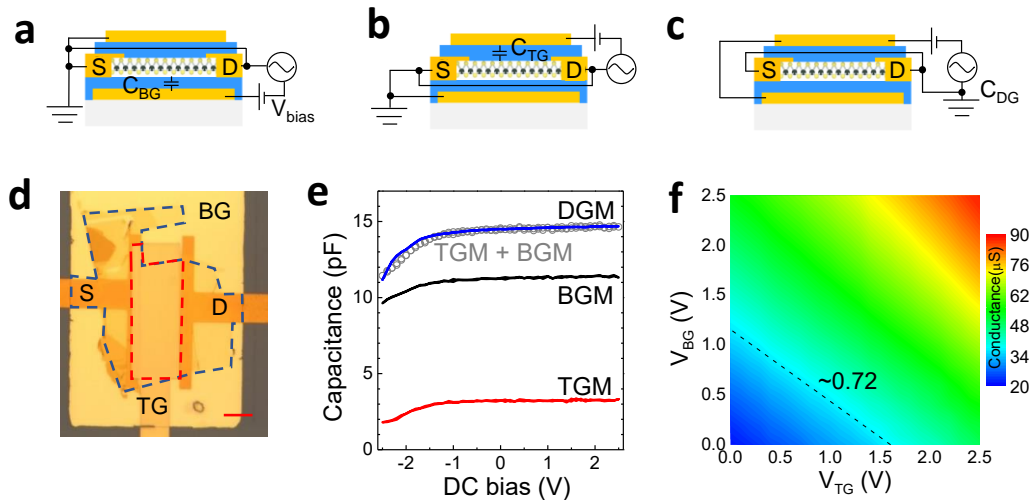

Figure S2. (a–c) Wiring geometry used in the C-V measurement for the dual-gated MoS$_2$ FET under BGM, TGM, and DGM. (d) Optical images for a large-area MoS$_2$ flake used for capacitive characterization. The contact areas of TG and BG are ~1780 μm$^2$ (dotted blue line) and ~710 μm$^2$ (dotted red line), respectively. Scale bar: 10 μm. (e) Measured capacitance values under different gating modes. (f) Contour plot for channel conductance as a function of $V_{TG}$ and $V_{BG}$.

## 4. Electrical data for 2L–4L FET channels

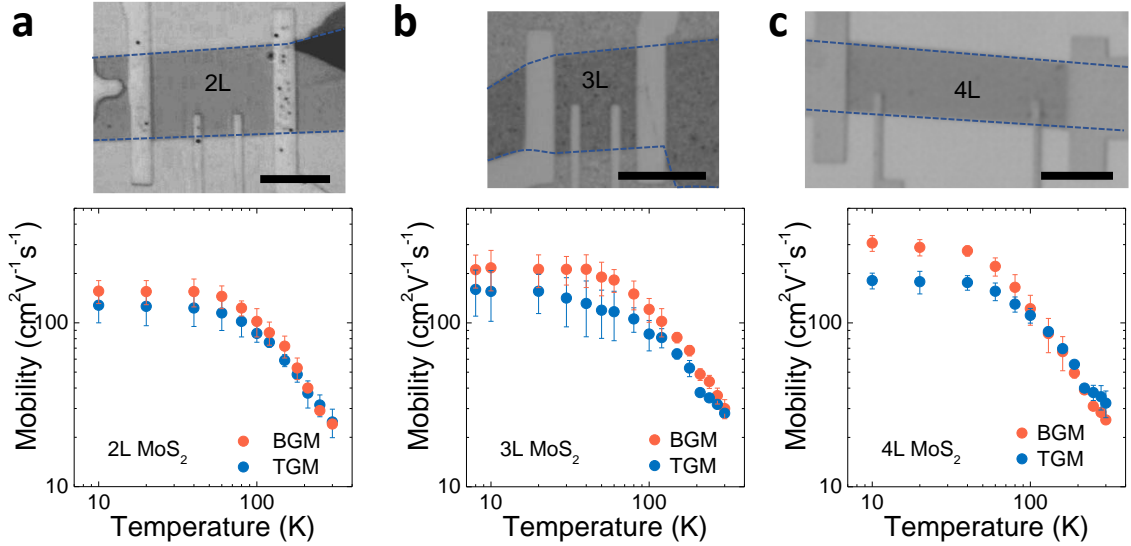

Figure S3. (a–c) Dependence of carrier mobility on $T$ for the 2L–4L MoS$_2$ FETs under BGM and TGM. To clearly show the employed channels, on the top of each panel are the device images just before depositing TG stacks. Scale bar: 10 μm.

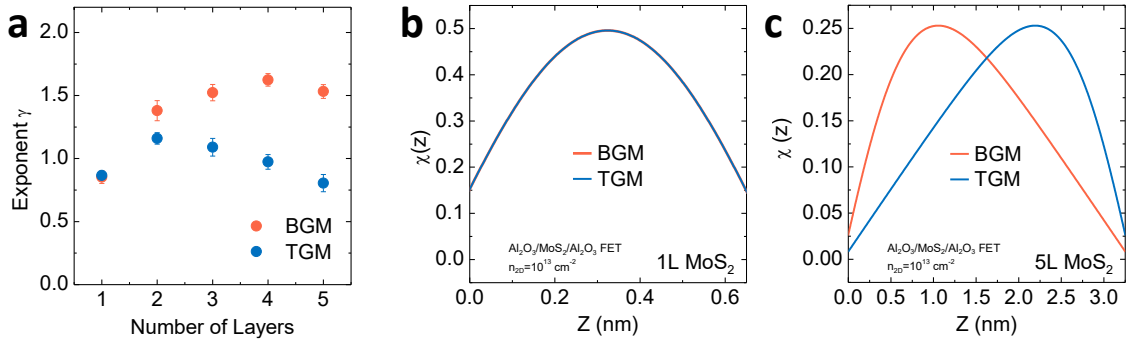

Figure S4. (a) Extracted $\gamma$ versus channel thickness under BGM and TGM, from the $\mu \propto T^{-\gamma}$ power-law fittings. (b, c) Comparison of carrier distributions under different gating modes and within varied channel thicknesses. In the atomically thin 1L channel, the distribution functions exhibit negligible difference between BGM and TGM due to the extremely reduced thickness, while the distribution functions change remarkably for the thick 5L channel. The left and right longitudinal axes in (b) and (c) represent the bottom and top interfaces of MoS$_2$ channels, respectively.

## 5. Theoretical calculations for various scattering mechanisms

**CI scattering**. According to Boltzmann's transport theory, the scattering rate from CIs is written as

$$\frac{1}{\tau_m(\mathbf{k})} = \frac{2\pi}{\hbar} n_i \sum_{\mathbf{q}} V_{\mathbf{q}}^2 (1 - \cos\theta) \delta(E_{\mathbf{k}} - E_{\mathbf{k}'}), \quad (A1)$$

where $n_i$ is the CI concentration, $q = 2k\sin(\theta/2)$ is the scattering vector with $\theta$ being the scattering angle from the initial momentum $\mathbf{k}$ to the final momentum $\mathbf{k}'$.

In equation (A1), the scattering matrix element can be written as

$$V_q = \frac{e^2 F_{CI}(q)}{2\varepsilon_s \varepsilon_{2D}(q) q}, \quad (A2)$$

where $F_{CI}(q)$ devotes the CI-related form factor, with its expression as

$$F_{CI}(q) = \frac{\varepsilon_1}{\varepsilon_1 + \varepsilon_2} \int_0^t \chi^2(z) e^{-q(t-z)} dz \sum_{n=0}^{\infty} \left(\xi e^{-2qt}\right)^n. \quad (A3)$$

In equation (A3), $t$ is the channel thickness, $\xi = \frac{\varepsilon_1 - \varepsilon_2}{\varepsilon_1 + \varepsilon_2} \frac{\varepsilon_1 - \varepsilon_3}{\varepsilon_1 + \varepsilon_3}$, and $\chi(z)$ is the carrier wave functions along the z direction, as given in Figure S4b,c, which can be achieved by jointly solving the Poisson and Schrodinger equations.

In equation (A2), the dielectric function of quantum screening is

$$\varepsilon_{2D}(\mathbf{q}) = 1 + \frac{e^2}{2\varepsilon_0 \varepsilon_s \mathbf{q}} F_{ee}(\mathbf{q}) \Pi(\mathbf{q}), \quad (A4)$$

where the 2D finite-temperature electron polarizability $\Pi(\mathbf{q}) = \frac{1}{S} \sum_{\mathbf{k}} \frac{f(E_{\mathbf{k}}) - f(E_{\mathbf{k+q}})}{E_{\mathbf{k+q}} - E_{\mathbf{k}}}$ with $f(E_{\mathbf{k}})$ being the Fermi-Dirac distribution function at the energy level of $E_{\mathbf{k}}$ and the overall polarization function

$$F_{ee}(q) = \left( F_{ee0}(q,t) + \frac{2\xi F_{ee1}(q,t) + \frac{\varepsilon_1 - \varepsilon_3}{\varepsilon_1 + \varepsilon_3} F_{ee2}(q,t) + \frac{\varepsilon_1 - \varepsilon_2}{\varepsilon_1 + \varepsilon_2} F_{ee3}(q,t)}{1 - \xi e^{-2qt}} \right). \quad (A5)$$

In our work, $\varepsilon_1 = \varepsilon_s$, $\varepsilon_2 = \varepsilon_3 = \varepsilon_{ox}$ and $F_{eei}(q,t)$ (i = 0, 1, 2, 3) represents the individual component of the form factor arising from the mirror forces of Coulomb interaction, and their expressions are

$$F_{ee0}(q,t) = \int_{-t/2}^{t/2} \int_{-t/2}^{t/2} \chi^2(z_1) \chi^2(z_2) e^{-q|z_1 - z_2|} dz_1 dz_2, \quad (A6)$$

$$F_{ee1}(q,t) = \int_{-t/2}^{t/2} \int_{-t/2}^{t/2} \chi^2(z_1)\chi^2(z_2) e^{-q(z_2-z_1)} dz_1 dz_2, \tag{A7}$$

$$F_{ee2}(q,t) = \int_{-t/2}^{t/2} \int_{-t/2}^{t/2} \chi^2(z_1)\chi^2(z_2) e^{-q(t-z_1-z_2)} dz_1 dz_2, \tag{A8}$$

$$F_{ee3}(q,t) = \int_{-t/2}^{t/2} \int_{-t/2}^{t/2} \chi^2(z_1)\chi^2(z_2) e^{-q(t+z_1+z_2)} dz_1 dz_2. \tag{A9}$$

**PLO phonons**. The scattering rate of PLO phonons is calculated by [2,3]

$$\frac{1}{\tau_m(\mathbf{k})} = \frac{2\pi}{\hbar} \sum_{\mathbf{q}} V_{LO}^2(\mathbf{q}) \left[1 - f\left(E_{\mathbf{k}+\mathbf{q}} \mp \hbar\omega_{LO}\right)\right] \delta\left(E_{\mathbf{k}+\mathbf{q}} - E_{\mathbf{k}} \mp \hbar\omega_{LO}\right), \tag{A10}$$

where 
$$V_{LO}^2(q) = \frac{e^2 L \hbar \omega_{LO}}{2 A_u \varepsilon_0} \left(N_{\omega_{LO}} + \frac{1}{2} \pm \frac{1}{2}\right) \left(\frac{1}{k_\infty} - \frac{1}{k_0}\right) F_{LO}^2(\mathbf{q}) \tag{A11}$$

and 
$$F_{LO}(q) = \int_0^t \frac{\left[1 - \exp(-tq/2)\cosh(qz)\right]}{Lq} \chi^2(z) dz. \tag{A12}$$

**ADP phonons**. The scattering rate of ADP from phonons is calculated by[2,3]

$$\frac{1}{\tau_m^{ADP}(k)} = \pi \left(\frac{m^* \Xi_1^2 k_B T}{\pi \hbar^3 \rho v_{s1}^2} + \frac{m^* \Xi_2^2 k_B T}{\pi \hbar^3 \rho v_{s1}^2}\right), \tag{A13}$$

**SR defects**. In the MoS$_2$ flakes, there is a high density of sulfur vacancies. They act as short-range scattering sources, and the corresponding scattering rate can be estimated by

$$\frac{1}{\tau_m(\mathbf{k})} = \frac{2\pi}{\hbar} n_{DE} \sum_{\mathbf{q}} V_q^2 (1 - \cos\theta) \delta(E_{\mathbf{k}} - E_{\mathbf{k}'}) \left[1 - f(E_{\mathbf{k}'})\right] \tag{A14}$$

where $n_{DE}$ is the defect concentration and $V_q = e^2 L / [2\varepsilon_s \varepsilon_{2D}(q)]$.

**SO phonons**. The surface optical phonons originate from the long-range Coulomb interactions between channel carriers and the excited phonons in the nearby dielectrics. Their strengths are associated with the characteristic frequencies of dielectrics. The scattering rate of SO phonons can be calculated by[3]

$$\frac{1}{\tau_m^{SO}} = \frac{2\pi}{\hbar} e^2 F_{SO} \sum_q \frac{N_{SO} + \frac{1}{2} \pm \frac{1}{2}}{q} \left[\frac{g(q)}{\varepsilon_{2D}(q)}\right]^2 (1 - \frac{\mathbf{k}+\mathbf{q}}{\mathbf{k}}\cos\theta) \delta(E_{\mathbf{k}+\mathbf{q}} - E_{\mathbf{k}} \pm \hbar\omega_{SO}) \tag{A15}$$

where the coupling strength of SO phonons is given by

$$F_{SO} = \frac{\hbar\omega_{SO}}{2A\varepsilon_0}\left(\frac{1}{\varepsilon_{ox}^\infty + \varepsilon_s^\infty} - \frac{1}{\varepsilon_{ox}^0 + \varepsilon_s^\infty}\right) \quad (A16)$$

$$g(q) = \int_0^t \chi^2(x)e^{-qx}dx \quad (A17)$$

where $A$ is surface area of the channel.

Table S1. Values of electrical permittivities adopted in calculation.

| MoS$_2$ thickness | 1 | 2 | 3 | 4 | 5 | bulk |
|---|---|---|---|---|---|---|
| $\varepsilon_s/\varepsilon_0$ | 4.8 | 6.85 | 8.02 | 8.78 | 9.33 | 12.8 |

Table S2. Parameters adopted in lattice phonons.[4]

| | |
|---|---|
| $m^*$ | $0.52m_0$ |
| $n_{2D}$ | $10^{13}$ cm$^{-2}$ |
| $t_s$ | 0.65nm (1L) |
| $\hbar\omega_{LO}$ | 48 meV |
| $\varepsilon_{ox}, \varepsilon_{ox}^\infty$ | $7\varepsilon_0, 3.2\varepsilon_0$ |
| $\Xi_1, \Xi_2$ | 1.6, 2.8 eV |
| $n_{DE}$ for 1L(5L) | 3 (6)×10$^{13}$ cm$^{-2}$ |
| $V_{s1}, V_{s2}$ | 4200, 6700 m/s |
| $k_\infty$ | $\varepsilon_s/(391.7/388.5)^2$ |
| $\rho$ | $3.1\times10^{-7}$ kg/m$^2$ |
| $\hbar\omega_{SO1}, \hbar\omega_{SO2}$ | 48.2meV, 71.4meV |

**References**
1. Li, S.-L.; Miyazaki, H.; Song, H.; Kuramochi, H.; Nakaharai, S.; Tsukagoshi, K. Quantitative Raman Spectrum and Reliable Thickness Identification for Atomic Layers on Insulating Substrates. *ACS Nano* **2012**, *6*, 7381–7388.
2. Li, S.-L.; Tsukagoshi, K.; Orgiu, E.; Samori, P. Charge Transport and Mobility Engineering in Two-Dimensional Transition Metal Chalcogenide Semiconductors. *Chem. Soc. Rev.* **2016**, *45*, 118–151.
3. Ma, N.; Jena, D. Charge Scattering and Mobility in Atomically Thin Semiconductors. *Phys. Rev. X* **2014**, *4*, 011043.
4. Kaasbjerg, K.; Thygesen, K. S.; Jacobsen, K. W. Phonon-Limited Mobility in n-Type Single-Layer MoS$_2$ from First Principles. *Phys. Rev. B* **2012**, *85*, 115317.